\documentclass[journal,onecolumn]{IEEEtran}
\usepackage{amsmath,amsfonts}
\usepackage{algorithmic}
\usepackage{algorithm}
\usepackage{array}
\usepackage{hyperref}
\usepackage[caption=false,font=normalsize,labelfont=sf,textfont=sf]{subfig}
\usepackage{textcomp}
\usepackage{stfloats}
\usepackage{url}
\usepackage{verbatim}
\usepackage{bm}
\usepackage{graphicx}
\usepackage{cite}
\usepackage{booktabs}
\usepackage{multirow}
\usepackage{xcolor}
\begin{document}

\title{Bayesian Joint Reconstruction and Uncertainty Quantification for Limited-Data Multi-Contrast X-Ray CT}

\author{Hassan Yazdanian and Henrik Mäkinen%
\thanks{This work was supported by the Research Council of Finland (RCF; Flagship of Advanced Mathematics for Sensing, Imaging and Modelling, grant 359186) and Business Finland through the TomoHQ project (funding decisions 591/31/2026 and 598/31/2026). \textit{Corresponding author: Hassan Yazdanian.}}%
\thanks{Hassan Yazdanian is with the Research Unit of Mathematical Sciences, University of Oulu, Oulu, Finland (e-mail: hassan.yazdanian@oulu.fi).}%
\thanks{Henrik Mäkinen is with the Department of Physics, University of Helsinki, Helsinki, Finland (e-mail: henrik.makinen@helsinki.fi).}%
}




\maketitle

\begin{abstract}
Multi-contrast X-ray computed tomography (CT) combines attenuation contrast with phase and dark-field contrast, which are sensitive to refraction and ultra-small-angle scattering, respectively. Common implementations record several intensity measurements per projection angle by phase or mask stepping, which lengthens acquisition and, at a fixed exposure per measurement, raises the dose. Limited-data acquisition via sparse angular sampling or reduced stepping lowers these costs but makes the data less informative, calling for prior information and uncertainty quantification. We study joint Bayesian reconstruction of attenuation, phase, and dark-field images from separately retrieved sinograms. A joint total variation term in the prior couples edge locations across the images but not their values. Posterior sampling with the No-U-Turn Sampler complements maximum a posteriori estimates with pixel-wise uncertainty. We compare this joint prior with channel-wise total variation alone on synthetic and experimental Talbot--Lau interferometry data under full, sparse-angle, sparse-step, and combined acquisition (up to 24 times fewer exposures). Both maximum a posteriori reconstructions suppressed undersampling artifacts, with synthetic relative errors 1.6--11 times lower than those of filtered backprojection. The joint prior gave modest synthetic accuracy gains, a cleaner experimental dark-field background, 23--30\,\% lower mean dark-field posterior standard deviation, and 4--38\,\% higher median effective sample size in experimental limited-data runs. In the synthetic study, posterior standard deviations matched or exceeded the errors in flat regions but underestimated them at boundaries, especially in phase. Experimental phase chains mixed slowly under angular undersampling. The formulation also applies to edge illumination with adapted forward and noise models.
\end{abstract}


\section{Introduction}
\IEEEPARstart{C}{onventional} X-ray computed tomography (CT) provides limited contrast between tissues and materials with similar attenuation properties \cite{bech2009soft,taba2018x,glinz2024comparison}. Phase-contrast and dark-field imaging provide complementary sensitivity to refraction and ultra-small-angle scattering, respectively \cite{pfeiffer2006phase,bech2010quantitative}. We refer to the combined use of attenuation, phase, and dark-field information for tomographic reconstruction as \emph{multi-contrast X-ray CT}.

Talbot--Lau interferometry (TLI) and edge illumination (EI) provide these contrasts using conventional X-ray sources \cite{pfeiffer2013grating,olivo2021edge}. Common implementations acquire multiple intensity measurements at each projection angle through phase stepping in TLI \cite{pfeiffer2006phase,bech2010quantitative} or mask stepping in EI \cite{olivo2021edge}. These measurements sample an intensity curve at each detector pixel. Relative to a reference scan without the object, attenuation lowers the mean of this curve, and refraction shifts it laterally, whereas ultra-small-angle scattering smears it out, which lowers its visibility (amplitude relative to the mean) in TLI and broadens it in EI \cite{bech2010quantitative,six2024joint}; the three signals can therefore be separated. The measured intensities can be reconstructed directly in the intensity domain \cite{brendel2016penalized,ritter2016simultaneous,xu2022intensity,six2024joint}, or by a two-step pipeline: signal retrieval first extracts attenuation, differential phase, and dark-field sinograms, which are then reconstructed by filtered backprojection (FBP) \cite{pfeiffer2007hard,bech2010quantitative,makinen2024optimization} or by iterative regularized methods \cite{xu2012investigation}.

The additional measurements increase acquisition time compared with conventional attenuation-based CT \cite{xu2022intensity,viermetz2022dark} and, at a fixed exposure per measurement, the dose; gratings or masks behind the sample further raise the dose needed for a given image quality, because they absorb photons that have already passed through it \cite{six2024joint,makinen2024optimization}. Sparse angular sampling \cite{bian2010evaluation,xu2012investigation} and reduced stepping \cite{ritter2016simultaneous,brendel2016penalized} can lower the acquisition time and the dose per scan; we refer to these strategies collectively as \emph{limited-data multi-contrast CT}. The two strategies degrade the data in different ways. Sparse angular sampling leaves the tomographic problem underdetermined, so that FBP produces streak artifacts \cite{hansen2021computed}, whereas reduced stepping at a fixed exposure per step records fewer photons and therefore increases the noise in the retrieved sinograms \cite{revol2010noise,weber2011}. Regularized iterative reconstruction suppresses streaks and noise by introducing prior information \cite{hansen2021computed}. The prior, however, also shapes features that the data constrain only weakly, and a regularized image alone does not reveal which of its features are supported by the measurements. Limited-data multi-contrast CT therefore raises two questions: what prior information should be imposed on the contrast images, and how the uncertainty that remains after regularization can be quantified.


For the first question, shared structure is a natural source of prior information: the contrast images describe different physical properties of the same object but may share boundaries despite differences in their amplitudes. Several joint reconstruction methods estimate all contrast channels directly from the measured intensities \cite{brendel2016penalized,ritter2016simultaneous,xu2022intensity,six2024joint,chen2017single}. Because each intensity depends on every channel, these methods couple the channels through the measurement model rather than through a prior; regularization, when used, is often applied to each channel separately \cite{brendel2016penalized,xu2022intensity,chen2017single,chen2018properties}. In EI simulations, fixing one channel to a noisy or blurred estimate had little influence on the reconstruction of the other \cite{chen2018properties}; likewise, in grating-based reconstructions, each channel's penalty mainly affected its own image \cite{brendel2016penalized}. These observations indicate robustness in the studied settings, but they do not show that shared structure is used to support a weaker channel.

Other joint reconstruction methods couple the channels explicitly. One penalizes differences between the phase image and a tissue-dependent mapping of the attenuation image \cite{patomontemayor2024intensity}, and another imposes such a material relationship exactly to reconstruct a single fused image \cite{vangogh2023towards}. These methods couple voxel values through a prescribed material model. Structural regularizers instead couple the locations or orientations of edges and leave the image values free. They have been used in joint positron emission tomography/magnetic resonance imaging \cite{ehrhardt2015joint} and multi-energy CT \cite{rigie2015joint,toivanen2020joint}, and for
attenuation and phase in grating-based radiography \cite{Nilchian16}. For multi-energy CT with fewer, non-overlapping projections per energy, a comparison of several joint regularizers, among them joint total variation (JTV), found that joint regularization outperformed channel-by-channel reconstruction \cite{toivanen2020joint}. Phase-guided structural coupling has recently been used for attenuation reconstruction in near-field holotomography \cite{liu2026joint}. These studies motivate priors that share spatial information across attenuation, phase, and dark-field images while allowing their values to differ.

Bayesian reconstruction addresses the second question: the posterior distribution of the unknown images, which combines the prior with the likelihood of the data, characterizes the uncertainty that remains under
the assumed models and complements point estimates with pixel-wise uncertainty maps \cite{kaipio2005statistical}. Previous CT studies have quantified uncertainty in reconstructed images \cite{bardsley2012mcmc,suuronen2020enhancing,horst2025uncertainty}, inclusion boundaries \cite{afkham2023uncertainty}, and suspected emboli \cite{rambojun2023uncertainty}, including under sparse-angle and limited-angle acquisition. Posterior uncertainty quantification (UQ) for jointly reconstructed attenuation, phase, and dark-field images remains comparatively little explored.

In this work, we study a joint Bayesian reconstruction for limited-data multi-contrast CT from separately retrieved sinograms. Because each retrieved sinogram depends on a single contrast image, the channels interact only through the prior when the channel noise is modeled as independent. The prior combines channel-wise total variation (TV) with a smoothed JTV term that couples edge locations across the three contrast images but not their values. Maximum a posteriori (MAP) estimates are complemented by posterior sampling with the No-U-Turn Sampler (NUTS) \cite{hoffman2014no}, which provides pixel-wise uncertainty. Using synthetic and experimental TLI data, we compare channel-wise and joint priors under four acquisition conditions (full data, sparse angles, sparse steps, and combined undersampling) and assess reconstruction accuracy, posterior variability (pixel-wise standard deviation), its calibration against actual error in the synthetic study, and sampling efficiency. These ingredients are established; our contribution is their combination for multicontrast TLI tomography and an assessment of what the joint prior and the posterior summaries deliver, and where they fall short.


\section{Methods}\label{sec:method}

We formulate multi-contrast CT reconstruction from retrieved sinograms as a Bayesian inverse problem, combining channel-wise forward and noise models with prior information to estimate the attenuation, phase, and dark-field images and quantify their uncertainty.

\subsection{Forward Problem}\label{subsec:forward}
The forward problem is to predict the noise-free attenuation, differential phase, and dark-field sinograms from given cross-sectional parameter images and a known acquisition geometry. Let \(\boldsymbol{\mu},\boldsymbol{\delta},\boldsymbol{\epsilon}\in\mathbb{R}^{N}\) denote the discretized attenuation coefficient, refractive-index decrement, and dark-field scattering coefficient on an image grid with \(N\) pixels. After signal retrieval, \(\boldsymbol{A},\boldsymbol{C},\boldsymbol{D}\in\mathbb{R}^{M}\) contain the logarithmic attenuation, differential phase-contrast (DPC), and logarithmic dark-field sinograms. The discrete line-integral projector is \(R\in\mathbb{R}^{M\times N}\), where \(M=N_\theta N_s\) for \(N_\theta\) projection angles and \(N_s\) detector positions. It approximates each line integral by quadrature along the ray, evaluating the image at the quadrature points by nearest-neighbor or bilinear interpolation. The forward maps for $\boldsymbol{A}$, $\boldsymbol{C}$, and $\boldsymbol{D}$ are
\begin{align}
\mathcal{F}_{\mu}(\boldsymbol{\mu}) &= R\boldsymbol{\mu},
\label{eq:attenuation_forward}\\
\mathcal{F}_{\delta}(\boldsymbol{\delta}) &= -G_sR\boldsymbol{\delta},
\label{eq:dpc_forward}\\
\mathcal{F}_{\epsilon}(\boldsymbol{\epsilon}) &= R\boldsymbol{\epsilon}.
\label{eq:darkfield_forward}
\end{align}
For DPC, \(G_s\in\mathbb{R}^{M\times M}\) is the detector-coordinate derivative in the interferometer's sensitivity direction, applied separately at each angle \cite{pfeiffer2006phase,vonteuffenbach2017grating}. The geometry-dependent constant that converts the refraction angle into the measured phase shift is absorbed into \(\boldsymbol{\delta}\), which is therefore reconstructed in scaled units; the sign convention depends on the setup. The dark-field model assumes that the visibility, like the intensity, decays exponentially along each ray \cite{wang2009quantitative,bech2010quantitative}.

Let \(\boldsymbol{x}\in\mathbb{R}^{3N}\) stack \((\boldsymbol{\mu},\boldsymbol{\delta},\boldsymbol{\epsilon})\), and let \(\boldsymbol{y}\in\mathbb{R}^{3M}\) stack \((\boldsymbol{A},\boldsymbol{C},\boldsymbol{D})\). The observation model is
\begin{equation}
\boldsymbol{y}=\mathcal{F}(\boldsymbol{x})+\boldsymbol{e},
\label{eq:deterministic_observation_model}
\end{equation}
where \(\mathcal{F}:\mathbb{R}^{3N}\to\mathbb{R}^{3M}\) stacks \(\mathcal{F}_{\mu}\), \(\mathcal{F}_{\delta}\), and \(\mathcal{F}_{\epsilon}\) in that order, and \(\boldsymbol{e}\) includes measurement and signal-retrieval errors.

\subsection{Bayesian Inverse Problem}

The inverse problem is to estimate the cross-sectional parameters \(\boldsymbol{x}\) from the noisy multi-contrast observations \(\boldsymbol{y}\) in \eqref{eq:deterministic_observation_model}. In the Bayesian formulation, the parameters and noise are modeled as random vectors \(\boldsymbol{X}\) and \(\boldsymbol{E}\), giving the random data \(\boldsymbol{Y}\) through
\begin{equation}
  \boldsymbol{Y}=\mathcal{F}(\boldsymbol{X})+\boldsymbol{E}.
  \label{eq:bayesian_observation_model}
\end{equation}
The observed data \(\boldsymbol{y}\) are a realization of \(\boldsymbol{Y}\).

Let \(\pi_0(\boldsymbol{x})\) denote the prior density of the parameters and \(\pi_{\boldsymbol{E}}(\boldsymbol{e})\) the noise density. Assuming \(\boldsymbol{E}\) is independent of \(\boldsymbol{X}\), the additive model \eqref{eq:bayesian_observation_model} gives the likelihood
\begin{equation}
  \pi(\boldsymbol{y}\mid\boldsymbol{x})
  =\pi_{\boldsymbol{E}}\!\left(\boldsymbol{y}-\mathcal{F}(\boldsymbol{x})\right).
  \label{eq:likelihood_from_noise}
\end{equation}
The Bayesian solution is the posterior distribution of the parameters conditioned on the observed data. By Bayes' formula, its density is \cite{kaipio2005statistical}
\begin{equation}
  \pi(\boldsymbol{x}\mid\boldsymbol{y})
  =\frac{\pi(\boldsymbol{y}\mid\boldsymbol{x})\pi_0(\boldsymbol{x})}
         {\pi(\boldsymbol{y})}
  \propto\pi(\boldsymbol{y}\mid\boldsymbol{x})\pi_0(\boldsymbol{x}),
  \label{eq:Bayes}
\end{equation}
where \(\pi(\boldsymbol{y})\) is the normalizing constant, also called the evidence. Below, we specify the likelihood and prior and describe the resulting posterior.

\subsubsection{Likelihood}
We model the blocks of \(\boldsymbol{E}\) for \(\boldsymbol{A}\), \(\boldsymbol{C}\), and \(\boldsymbol{D}\) as mutually independent Gaussian random vectors, \(\boldsymbol{E}^{c}\sim\mathcal{N}(0,\sigma_c^2 I_M)\), \(c\in\{A,C,D\}\), where \(I_M\) is the identity matrix and \(\sigma_c^2\) is the corresponding channel variance, treated as known (Section~\ref{sec:experiments}). The likelihood is
\begin{equation}
\pi(\boldsymbol{y}\mid\boldsymbol{x})
 \propto\exp\!\Biggl(
-\frac{1}{2\sigma_A^2}
  \left\|\boldsymbol{A}-\mathcal{F}_{\mu}(\boldsymbol{\mu})\right\|_2^2
-\frac{1}{2\sigma_C^2}
  \left\|\boldsymbol{C}-\mathcal{F}_{\delta}(\boldsymbol{\delta})\right\|_2^2
-\frac{1}{2\sigma_D^2}
  \left\|\boldsymbol{D}-\mathcal{F}_{\epsilon}(\boldsymbol{\epsilon})\right\|_2^2
\Biggr),
\label{eq:likelihood}
\end{equation}
with \(\|\cdot\|_2\) denoting the Euclidean norm.



\subsubsection{Prior}
For \(\pi_0\), we consider two TV-based edge-preserving priors \cite{hansen2021computed,kolehmainen1998bayesian}. TV priors are not discretization-invariant \cite{lassas2004can}, so we use them at a fixed image resolution.  For \(\boldsymbol{u}\in\mathbb{R}^{N}\), represented on an \(n_1\times n_2\) grid with \(N=n_1n_2\), the smoothed isotropic TV functional is
\begin{equation}
\mathrm{TV}(\boldsymbol{u})=
\sum_{i=1}^{n_1-1}\sum_{j=1}^{n_2-1}
\sqrt{|\nabla\boldsymbol{u}_{i,j}|^2+\beta^2}.
\label{eq:smoothed_tv}
\end{equation}
Here \(\nabla\boldsymbol{u}_{i,j}\in\mathbb{R}^2\) contains the forward differences of \(\boldsymbol{u}\) at pixel \((i,j)\) in the two grid directions, and the sum runs over the pixels at which both differences are defined. The smoothing parameter \(\beta>0\) makes TV differentiable for optimization and sampling.

Because the channels differ in physical units and magnitude, the priors act on the scaled images \(\boldsymbol{\mu}/s_\mu\), \(\boldsymbol{\delta}/s_\delta\), and \(\boldsymbol{\epsilon}/s_\epsilon\), with fixed channel scales \(s_c>0\) \(c\in\{\mu,\delta,\epsilon\}\) given in Section~\ref{sec:experiments}; for brevity, the scales are omitted below. The channel-wise baseline prior is
\begin{equation}
\pi_{\mathrm{cw}}(\boldsymbol{x})\propto\exp\!\Bigl[
-\alpha_{\mu}\,\mathrm{TV}(\boldsymbol{\mu})
 -\alpha_{\delta}\,\mathrm{TV}(\boldsymbol{\delta})
-\alpha_{\epsilon}\,\mathrm{TV}(\boldsymbol{\epsilon})
 -\frac{\lambda_0}{2}\|\boldsymbol{x}\|_2^2
\Bigr],
\label{eq:channelwise_prior_density}
\end{equation}
where the channel weights \(\alpha_{\mu},\alpha_{\delta},\alpha_{\epsilon}>0\) set the strength of the TV term in each channel. The TV penalties are invariant to constant shifts and therefore do not define a proper density on their own \cite{lin2025proper}; the amplitude penalty with weight \(\lambda_0>0\) bounds the density by a Gaussian, so the prior is proper on \(\mathbb{R}^{3N}\).

Structural coupling is introduced through
\begin{equation}
\mathrm{JTV}(\boldsymbol{x})=
\sum_{i=1}^{n_1-1}\sum_{j=1}^{n_2-1}
\sqrt{
|\nabla\boldsymbol{\mu}_{i,j}|^2+
|\nabla\boldsymbol{\delta}_{i,j}|^2+
|\nabla\boldsymbol{\epsilon}_{i,j}|^2+\beta^2
},
\label{eq:joint_tv}
\end{equation}
which groups the channel gradients at each pixel. Because the Euclidean norm of the stacked gradients never exceeds the sum of the individual gradient magnitudes, edges that coincide across channels cost less than the same edges placed at different pixels. JTV therefore favors co-located edges without tying the size or sign of the jumps across channels. The joint prior is
\begin{equation}
\pi_{\mathrm{jtv}}(\boldsymbol{x})\propto
\pi_{\mathrm{cw}}(\boldsymbol{x})
\exp\!\left[-\alpha_{\mathrm{J}}\,\mathrm{JTV}(\boldsymbol{x})\right],
\label{eq:jtv_prior_density}
\end{equation}
where \(\alpha_{\mathrm{J}}>0\) is the coupling weight; \(\alpha_{\mathrm{J}}=0\) recovers \(\pi_{\mathrm{cw}}\).

The channel weights \(\alpha_\mu,\alpha_\delta,\alpha_\epsilon\), the coupling weight \(\alpha_{\mathrm{J}}\), the amplitude weight \(\lambda_0\), and the smoothing parameter \(\beta\) are referred to collectively as the \emph{prior parameters}.

\subsubsection{Posterior}
For \(q\in\{\mathrm{cw},\mathrm{jtv}\}\), let \(\pi_q(\boldsymbol{x}\mid\boldsymbol{y})\) denote the posterior obtained by setting \(\pi_0=\pi_q\) in \eqref{eq:Bayes}.
Up to an additive constant independent of \(\boldsymbol{x}\), the negative log-posterior is
\begin{equation}
\Phi_q(\boldsymbol{x})=
-\log\pi(\boldsymbol{y}\mid\boldsymbol{x})+
\mathcal{R}_q(\boldsymbol{x}),
\label{eq:negative_log_posterior}
\end{equation}
where \(\mathcal{R}_q\) is the negative log-prior. The likelihood \eqref{eq:likelihood} and $\pi_\mathrm{cw}$ factorize over channels, so only the JTV term couples the channels. Since the forward maps are linear and the smoothed TV and JTV terms are convex, the quadratic amplitude penalty makes \(\Phi_q\) strongly convex, so both posteriors are log-concave with a unique mode.



\subsection{Inference}

\subsubsection{MAP Estimate}
The MAP estimate minimizes the negative log-posterior:
\begin{equation}
\boldsymbol{x}_{\mathrm{MAP}}^{q}=
\arg\min_{\boldsymbol{x}\in\mathbb{R}^{3N}}\Phi_q(\boldsymbol{x}).
\label{eq:map_definition_min}
\end{equation}
This is a variational regularization problem with regularization functional \(\mathcal{R}_q\) \cite{kaipio2005statistical}, so the MAP estimate is also a regularized reconstruction and the prior weights act as regularization parameters.

\subsubsection{Sampling and UQ}
Posterior samples are generated using the No-U-Turn Sampler (NUTS) \cite{hoffman2014no}, a gradient-based Hamiltonian Monte Carlo method previously applied to single-channel CT \cite{suuronen2020enhancing}. The samples \(\boldsymbol{x}^{(k)}\), \(k=1,\ldots,K\), are the post-warmup draws of several independent NUTS chains. We estimate the posterior mean and pixel-wise standard deviation by
\begin{align}
\bar{\boldsymbol{x}}&=\frac{1}{K}\sum_{k=1}^{K}\boldsymbol{x}^{(k)},
\label{eq:posterior_mean}\\
\boldsymbol{s}&=\left[
\frac{1}{K-1}\sum_{k=1}^{K}
\left(\boldsymbol{x}^{(k)}-\bar{\boldsymbol{x}}\right)^2
\right]^{1/2},
\label{eq:posterior_std}
\end{align}
where powers are evaluated component-wise. The map $\boldsymbol s$ quantifies posterior variability conditional on the assumed forward model, noise model, and prior. We assess convergence with the Gelman--Rubin statistic \(\hat{R}\) and sampling efficiency with the effective sample size (ESS) \cite{mcbook}, both computed per pixel. Optimization and sampling settings for the synthetic and experimental datasets are specified in Section~\ref{sec:experiments}.

\section{Numerical Experiments and Results}\label{sec:experiments}

We compare FBP with the channel-wise TV \eqref{eq:channelwise_prior_density} and JTV \eqref{eq:jtv_prior_density} priors using synthetic data with known ground truth and experimental TLI data. Each dataset is tested with full data, sparse angles, sparse steps, and combined undersampling. We assess MAP reconstructions and posterior UQ against the ground truth for synthetic data; experimental evaluation uses data consistency and, as surrogate references, full-data reconstructions.

All reconstructions and posterior sampling ran on an NVIDIA RTX 6000 Ada Generation GPU (48~GB) under Ubuntu 22.04.5 LTS, using Python~3.14.2, NumPy~2.4.2, scikit-image 0.26.0 \cite{van2014scikit}, PyTorch~2.10.0 \cite{paszke2019pytorch} with CUDA~13.0, ASTRA Toolbox~2.4.1 \cite{van2016fast}, and Pyro~1.9.1 \cite{bingham2019pyro}. The code, including the experimental data used here, is available at \url{https://github.com/hassanyazdanian/Bayes_mc_CT}.


\subsection{Synthetic Phantom}\label{sec:exp:synthetic}
\subsubsection{Phantom and Data Simulation}\label{sec:exp:synthetic:setup}

\begin{figure}[!t]
\centering
\includegraphics[width=0.9\columnwidth]{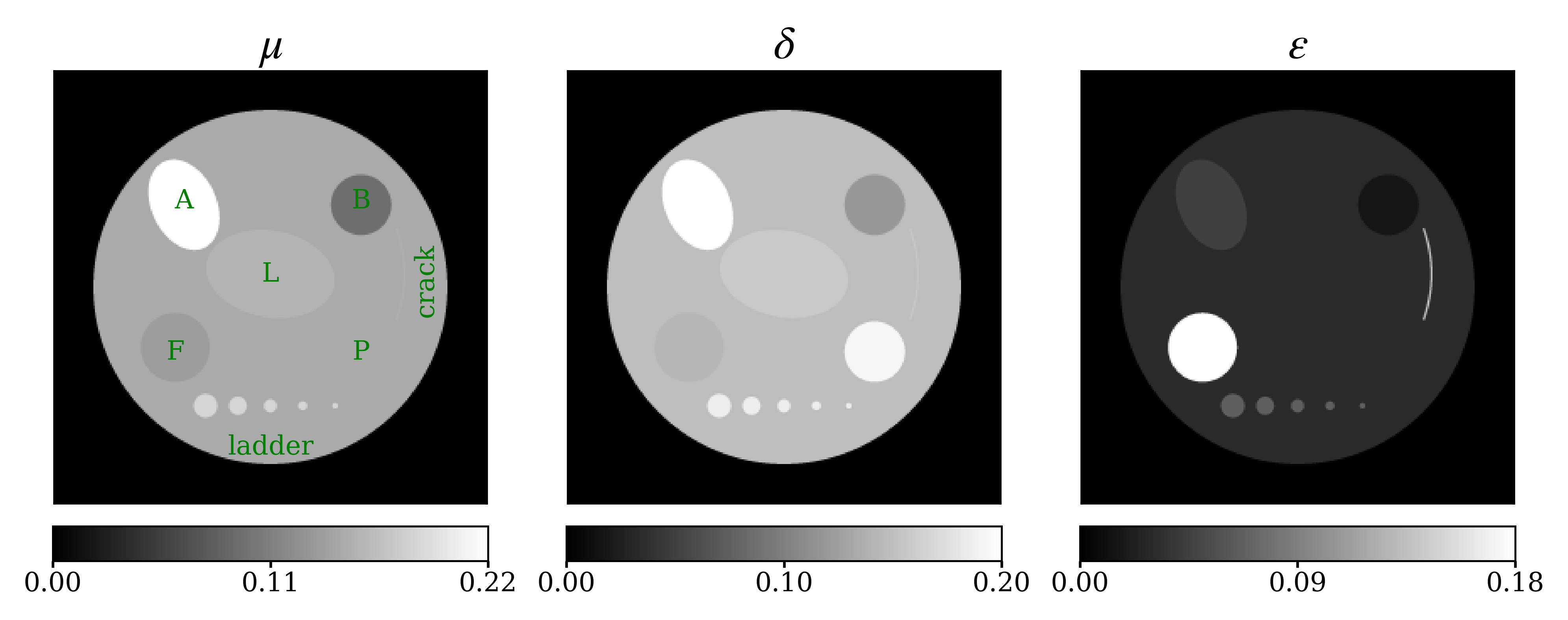}
\caption{Synthetic multi-contrast phantom: ground-truth attenuation \(\mu\), phase \(\delta\), and dark-field \(\epsilon\) images. Inclusions A and B share their boundaries across the channels, F is $\epsilon$-dominant, and P differs from the disc only in $\delta$; the low-contrast ellipse L, the ladder of decreasing-size discs, and the thin $\epsilon$-dominant crack serve the UQ study.}
\label{fig:synth:phantom}
\end{figure}

The synthetic phantom is a piecewise-constant disc in normalized units (Fig.~\ref{fig:synth:phantom}). Inclusions A and B share boundaries across channels, with contrast changes relative to the disc of $+50/{+35}/{+50}\,\%$ and $-35/{-20}/{-50}\,\%$ in $\mu/\delta/\epsilon$, respectively. They test coupling of edge locations across different contrast amplitudes. Cross-talk is tested by a foam-like inclusion F, weak in $\mu$ and $\delta$ ($-8$ and $-4\,\%$) but dominant in $\epsilon$ ($+500\,\%$), and an attenuation-matched inclusion P visible only in $\delta$. A ladder of five decreasing-size discs, a low-contrast ellipse L ($+6\,\%$ in $\mu$ and $\delta$, absent in $\epsilon$), and a one-pixel-wide arc-shaped crack, strong only in $\epsilon$, provide small or low-contrast features for the UQ study.

Phase-stepping intensities are simulated with the Talbot--Lau model at visibility $v=0.3$ and processed by the harmonic analysis used for the experimental data (see Sections~\ref{sec:exp:real:preproc} and~\ref{sec:exp:real:validation}). To avoid an inverse crime, the phantom is rendered on a $1024^2$ grid and block-averaged to $512^2$ for simulation with a bilinear-interpolation projector. Ground truth is block-averaged to the $256^2$ reconstruction grid, where a nearest-neighbor projector is used. All
three images are scaled by one common factor so that $\max|\mathrm{DPC}|\le0.8\pi$, below the phase-wrapping limit of $\pi$. Full data use $180$ views over $360^\circ$ and $10$ phase steps; the sparse settings use $30$ views and $4$ steps (the experimental study uses $5$ steps, since only divisors of its $10$ recorded steps give evenly spaced subsets).

Independent zero-mean Gaussian noise is added to the retrieved attenuation, DPC, and dark-field sinograms at $1\,\%$, $4.7\,\%$, and $4.8\,\%$ of their noiseless root-mean-square (RMS) values. The DPC and dark-field levels are the attenuation level times the phase-stepping noise factors $\sqrt{2}/v$ and $\sqrt{1+2/v^2}$ \cite{revol2010noise,weber2011,chabior2011signal}, which are derived for absolute noise levels and applied here to relative levels because the phantom is in normalized units. For $n_{\mathrm{phase}}$ steps, the levels are scaled by $\sqrt{10/n_{\mathrm{phase}}}$. On full data, the RMS difference between the noiseless simulated sinograms and the reconstruction model applied to the ground truth is approximately $1.2$, $5.1$, and $0.4$ noise standard deviations in attenuation, DPC, and dark-field, respectively; differentiation amplifies discretization differences in DPC.


\subsubsection{Reconstruction and Evaluation Settings}\label{sec:exp:synthetic:settings}
The synthetic fan-beam geometry has source-to-object and source-to-detector distances of $0.20$ and $0.40$\,m, with $256$ detector elements spanning $120$\,mm. Reconstruction uses a $256^2$ grid over a $60$\,mm field of view and $256\sqrt{2}$ quadrature points per ray. DPC follows \eqref{eq:dpc_forward} with second-order central differences, and the likelihood uses the standard deviations of the added noise. The priors act on the unscaled images (\(s_c=1\)), whose channels have comparable magnitudes. Bayesian inference is restricted to a circular support of radius $0.98$ in normalized image coordinates, in which the reconstruction grid spans $[-1,1]^2$. Outside this support, the images are fixed at zero. Fan-beam FBP uses a Ram--Lak filter on the full square grid, with integrated phase obtained by Fourier integration of the noisy DPC.

MAP estimates use L-BFGS with a strong-Wolfe line search \cite{nocedal2006numerical}, at most $1000$ iterations, and zero initialization for both priors. The same prior parameters $(\beta, \alpha_\mu,\alpha_\delta,\alpha_\epsilon, \lambda_0, \alpha_{\mathrm{J}})$ are used in all acquisition conditions. A joint sweep of $\beta$ and the channel weights on full data gave similar errors for $\beta\in[10^{-4},10^{-3}]$ (within $3.4\,\%$ of each channel's best value). We use $\beta=3\times10^{-4}$ to avoid the dark-field background texture at $10^{-3}$ and the stiffer posterior, which is harder to sample, at smaller values. Minimizing the relative $\ell_2$ error (RE) averaged over all four conditions gives $(\alpha_\mu,\alpha_\delta,\alpha_\epsilon)=(100,316,56.2)$, within $4\,\%$ of each per-condition optimum. Tuning on the full data alone would over-regularize the undersampled cases: the full-data optimum $\alpha_\delta=1778$ more than doubles the combined-undersampling phase error. The amplitude weight $\lambda_0=1$ is the largest tested value that changes RE by less than $0.3\,\%$ in every channel and condition. The coupling weight $\alpha_\mathrm J=29.6$ minimizes the average change in summed per-channel RE relative to TV, reducing that sum in every condition. Tuning against ground truth on the evaluated noise realization gives best-case performance for each prior.


Posterior samples are drawn with NUTS in normalized coordinates, each channel divided by the $99$th percentile of its MAP magnitude. Two chains of $1000$ samples are drawn after $500$ warmup iterations, during which the step size (initial value $10^{-2}$) and a diagonal mass matrix are adapted, with a maximum tree depth of six. Each chain starts from the MAP estimate of the corresponding prior with an independent perturbation of standard deviation $10^{-3}$. The sampling dimension is $147\,192$, i.e.\ $49\,064$ active pixels per channel.


Reconstructions are scored per channel against the ground truth by the RE, computed on the full image, and by the structural similarity (SSIM) index  \cite{wang2004image} and the peak signal-to-noise ratio (PSNR). For SSIM and PSNR, the true and reconstructed images are mapped by the same affine transformation, taking the range of the true image to $[0,1]$, and the reconstruction is clipped to that interval; both indices use the scikit-image \cite{van2014scikit} implementation at its default settings with unit data range.

\subsubsection{MAP Reconstructions}\label{sec:exp:synthetic:map}

\begin{figure}[!t]
\centering
\includegraphics[width=\columnwidth]{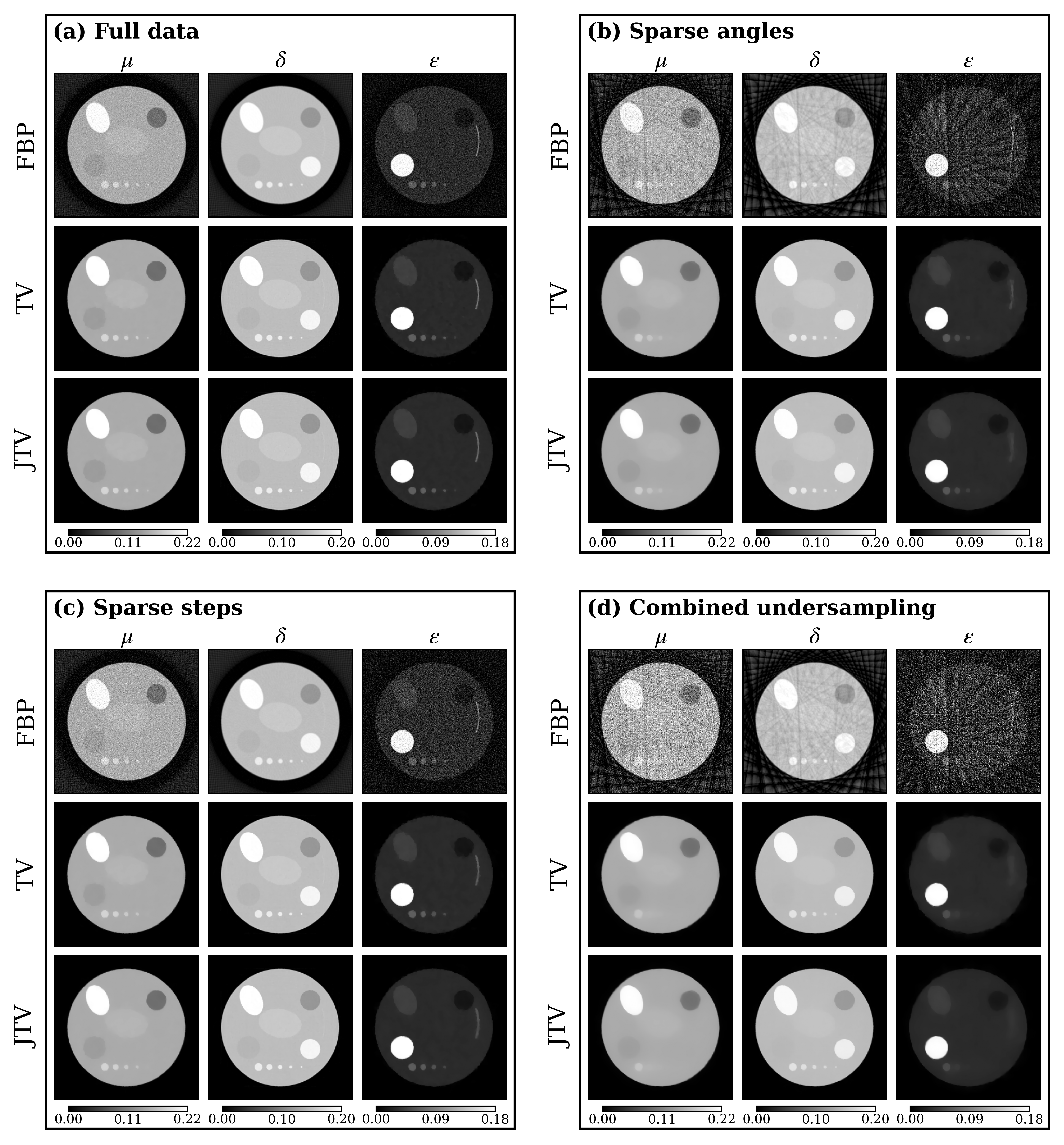}
\caption{Synthetic-data reconstructions of the attenuation ($\mu$), phase ($\delta$), and dark-field ($\epsilon$) channels under different acquisition conditions. (a) Full data: 180 projection angles and 10 phase steps; (b) sparse angles: 30 projection angles and 10 phase steps; (c) sparse steps: 180 projection angles and 4 phase steps; and (d) combined undersampling: 30 projection angles and 4 phase steps. For each acquisition condition, the rows show FBP, channel-wise TV, and JTV reconstructions, respectively. Within each panel, the three reconstructions of a given contrast are displayed using the same intensity scale, indicated by the corresponding colorbar.}
\label{fig:synth:map}
\end{figure}

Figure~\ref{fig:synth:map} and Table~\ref{tab:synth_metrics} compare the reconstructions. Under combined undersampling, the RE of FBP reaches $0.425$ in $\mu$ and $1.56$ in $\epsilon$; the phase channel remains more robust after Fourier integration ($0.170$). The Bayesian RE is $0.066$--$0.071$ in $\mu$ and $\delta$ and $0.111$--$0.146$ in $\epsilon$, $1.6$--$11$ times lower than FBP. For FBP and for the Bayesian dark-field channel, reducing the number of views degrades the reconstruction more than reducing the phase steps, whereas the Bayesian $\mu$ and $\delta$ errors vary by less than $5\,\%$ across conditions.



JTV lowers RE in $11$ of the $12$ channel--condition pairs; the exception is $\delta$ under combined undersampling ($+0.4\,\%$). Gains are modest, reaching $3.1\,\%$ in full-data $\epsilon$, with both priors tuned near their individual optima. Dark-field SSIM increases by $0.008$--$0.013$ in every condition, consistent with improved low-contrast boundaries. JTV also does not transfer channel-specific features: the magnitude of the contrast error of P in $\mu$ and $\epsilon$ and of F in $\mu$ and $\delta$, measured against the surrounding disc, stays within $0.01$ percentage points of TV or decreases, by up to $0.25$ points.

MAP estimation took $3$--$8$\,s per reconstruction on the GPU, against $2$--$5$\,ms for FBP.

\begin{table*}[!t]
\centering
\caption{Synthetic-data reconstruction accuracy against the ground truth. The better of TV and JTV is shown in bold where the displayed values differ. TV and JTV use at most $1000$ L-BFGS iterations.}
\label{tab:synth_metrics}

\footnotesize
\setlength{\tabcolsep}{5.6pt}
\renewcommand{\arraystretch}{0.95}

\begin{tabular}{cc ccc ccc ccc ccc}
\toprule
& &
\multicolumn{3}{c}{\textbf{(a) Full data}} &
\multicolumn{3}{c}{\textbf{(b) Sparse angles}} &
\multicolumn{3}{c}{\textbf{(c) Sparse steps}} &
\multicolumn{3}{c}{\textbf{(d) Combined}} \\
\cmidrule(lr){3-5}
\cmidrule(lr){6-8}
\cmidrule(lr){9-11}
\cmidrule(lr){12-14}

Channel & Method
& RE $\downarrow$ & SSIM $\uparrow$ & PSNR $\uparrow$
& RE $\downarrow$ & SSIM $\uparrow$ & PSNR $\uparrow$
& RE $\downarrow$ & SSIM $\uparrow$ & PSNR $\uparrow$
& RE $\downarrow$ & SSIM $\uparrow$ & PSNR $\uparrow$ \\
& & & & (dB) & & & (dB) & & & (dB) & & & (dB) \\
\midrule

\multirow{3}{*}{$\mu$}
& FBP
& 0.1472 & 0.265 & 23.03
& 0.3066 & 0.114 & 17.41
& 0.1903 & 0.174 & 20.96
& 0.4253 & 0.079 & 14.71 \\

& TV
& 0.0673 & 0.963 & 29.52
& 0.0677 & 0.948 & 29.46
& 0.0678 & 0.958 & 29.45
& 0.0705 & \textbf{0.923} & 29.11 \\

& JTV
& \textbf{0.0669} & \textbf{0.964} & \textbf{29.57}
& \textbf{0.0674} & 0.948 & \textbf{29.51}
& \textbf{0.0674} & \textbf{0.959} & \textbf{29.50}
& \textbf{0.0700} & 0.921 & \textbf{29.17} \\

\cmidrule(lr){2-14}

\multirow{3}{*}{$\delta$}
& FBP
& 0.1097 & 0.740 & 24.32
& 0.1687 & 0.432 & 21.61
& 0.1100 & 0.728 & 24.30
& 0.1695 & 0.419 & 21.57 \\

& TV
& 0.0676 & 0.937 & 28.59
& 0.0656 & 0.957 & 28.80
& 0.0666 & 0.954 & 28.68
& \textbf{0.0672} & 0.953 & \textbf{28.58} \\

& JTV
& \textbf{0.0674} & \textbf{0.940} & \textbf{28.60}
& \textbf{0.0655} & \textbf{0.958} & \textbf{28.81}
& \textbf{0.0665} & \textbf{0.955} & \textbf{28.70}
& 0.0675 & 0.953 & 28.54 \\

\cmidrule(lr){2-14}

\multirow{3}{*}{$\epsilon$}
& FBP
& 0.4110 & 0.156 & 23.16
& 1.0385 & 0.051 & 16.15
& 0.6325 & 0.092 & 19.82
& 1.5636 & 0.030 & 13.02 \\

& TV
& 0.1147 & 0.949 & 33.34
& 0.1284 & 0.933 & 32.32
& 0.1177 & 0.950 & 33.09
& 0.1456 & 0.904 & 31.23 \\

& JTV
& \textbf{0.1111} & \textbf{0.961} & \textbf{33.59}
& \textbf{0.1282} & \textbf{0.941} & \textbf{32.33}
& \textbf{0.1176} & \textbf{0.958} & 33.09
& \textbf{0.1429} & \textbf{0.917} & \textbf{31.39} \\

\bottomrule
\end{tabular}
\end{table*}

\subsubsection{Posterior Sampling and UQ}
\label{sec:exp:synthetic:uq}

Diagnostics over $2000$ randomly selected coordinates gave a maximum $\hat{R}$ of $1.02$--$1.04$ for $180$ views and $1.10$--$1.17$ for $30$ views, with median $\hat{R}\le1.004$. Median ESS values were $480$--$745$ and $143$--$189$, respectively, with minima of $142$--$311$ and $5$--$13$. Thus, some sparse-angle coordinates remained slowly mixing. Two-chain runs took $19$\,min at $30$ views and $94$\,min at $180$ views ($0.39$ and $1.89$\,s per iteration).

\begin{figure*}[!t]
    \centering
    \includegraphics[width=0.93\textwidth]{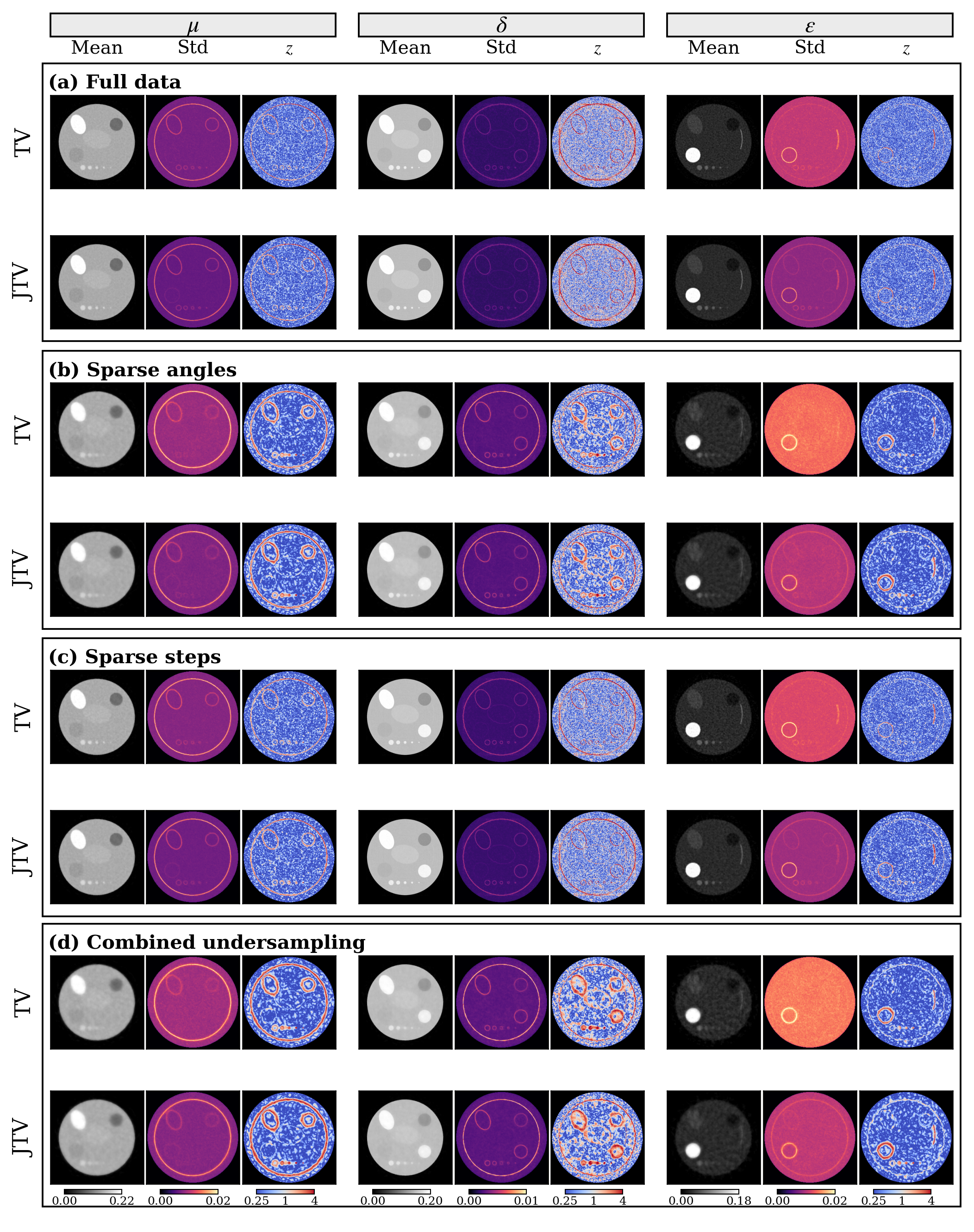}
    \caption{Synthetic-data posterior UQ for the four acquisition scenarios: (a) full data, (b) sparse angles, (c) sparse steps, and (d) combined undersampling. For each scenario, TV and JTV are compared in terms of the posterior mean, the pixel-wise posterior standard deviation $\boldsymbol{s}$, and the absolute standardized error $\boldsymbol{z}$, i.e.\ the posterior-mean error in units of $\boldsymbol{s}$ (Section~\ref{sec:exp:synthetic:uq}), for the attenuation $\mu$, phase $\delta$, and dark-field $\epsilon$ channels. Mean and standard deviation use a common scale per channel across all scenarios and priors; $z$ uses a logarithmic scale centered on the calibrated value $z=1$, with blue indicating a posterior wider than its error and red indicating overconfidence, saturating at $0.25$ and $4$. The colorbars at the foot of the figure apply to all panels.}
    \label{fig:uq_comparison}
\end{figure*}

Figure~\ref{fig:uq_comparison} shows posterior means $\bar{\boldsymbol{x}}$, standard deviations $\boldsymbol{s}$, and absolute standardized errors $\boldsymbol{z}=|\bar{\boldsymbol{x}}-\boldsymbol{x}_{\mathrm{true}}|/\boldsymbol{s}$, computed pixel-wise, for each channel. From full data to combined undersampling, the mean posterior standard deviation of the TV posterior increases by $30\,\%$, $57\,\%$, and $36\,\%$ in $\mu$, $\delta$, and $\epsilon$, with uncertainty concentrated near boundaries. Relative to TV, JTV reduces mean standard deviation by $12$--$14\,\%$ in $\mu$, about $2\,\%$ in $\delta$, and $23$--$27\,\%$ in $\epsilon$. For full-data $\epsilon$, JTV lowers the posterior-mean RE from $0.160$ to $0.138$, compared with $0.115$ to $0.111$ for the MAP estimate. The posterior mean remains less accurate than MAP in $\mu$ and $\epsilon$, while for $\delta$ the two differ by at most $1.1\,\%$.


Table~\ref{tab:synth:uq} reports the RMS error of the posterior mean divided by the RMS posterior standard deviation, on edge pixels 
and on the remaining flat pixels. In flat regions, the ratio is $0.37$--$0.74$ for $\mu$ and $\epsilon$, with $90$--$99\,\%$ of pixels within one standard deviation, and $0.79$--$1.02$ for $\delta$, with $73$--$83\,\%$. At edges, the ratios are $2.5$--$3.2$ for $\mu$, $5.3$--$10.5$ for $\delta$, and $1.2$--$1.8$ for $\epsilon$; two-standard-deviation coverage falls to $35$--$81\,\%$ in $\mu$ and $49$--$65\,\%$ in $\delta$. Against Gaussian reference coverages of $68\,\%$ and $95\,\%$, these results indicate conservative uncertainty for $\mu$ and $\epsilon$ in flat regions, roughly matched uncertainty for $\delta$ there, and underestimated errors at boundaries, also visible in the standardized error maps.


For $\delta$, the edge-error ratio decreases from $10.3$ to $5.3$ as data are removed, while the edge error stays near $0.026$: posterior variability grows while systematic error persists. Reconstructing with the projector used for data simulation (Section~\ref{sec:exp:synthetic:setup}) reduces the edge error to $0.016$, about $40\,\%$ lower. This points to model discrepancy and grid-representation errors that the assumed-exact forward model does not include in the posterior. These are per-pixel marginal diagnostics and do not assess joint posterior coverage.

\begin{table}[!t]
  \centering
  \caption{Calibration of the synthetic posteriors: RMS error of the posterior mean divided by RMS posterior standard deviation, for pixels on true edges and in flat regions. A calibrated posterior gives one; larger values indicate overconfidence.}
  \label{tab:synth:uq}
  \footnotesize
  \begin{tabular}{llcccccc}
    \toprule
     & & \multicolumn{2}{c}{$\mu$} & \multicolumn{2}{c}{$\delta$} & \multicolumn{2}{c}{$\epsilon$} \\
    \cmidrule(lr){3-4}\cmidrule(lr){5-6}\cmidrule(lr){7-8}
    Scenario & Prior & flat & edge & flat & edge & flat & edge \\
    \midrule
    Full data     & TV  & 0.45 & 2.92 & 1.01 & 10.33 & 0.47 & 1.30 \\
                  & JTV & 0.44 & 3.21 & 1.02 & 10.54 & 0.45 & 1.51 \\
    \midrule
    Sparse angles & TV  & 0.51 & 2.50 & 0.79 & 5.74  & 0.37 & 1.27 \\
                  & JTV & 0.56 & 2.82 & 0.82 & 5.90  & 0.42 & 1.63 \\
    \midrule
    Sparse steps  & TV  & 0.43 & 2.58 & 0.79 & 8.76  & 0.44 & 1.21 \\
                  & JTV & 0.43 & 2.89 & 0.79 & 8.92  & 0.42 & 1.48 \\
    \midrule
    Combined      & TV  & 0.62 & 2.77 & 0.92 & 5.34  & 0.38 & 1.35 \\
                  & JTV & 0.74 & 3.10 & 0.95 & 5.48  & 0.46 & 1.76 \\
    \bottomrule
  \end{tabular}
\end{table}

\subsection{Real Data: Talbot--Lau Interferometry}\label{sec:exp:real}

\subsubsection{TLI Setup and Phantom}\label{sec:exp:real:setup}

\begin{figure}[!t]
\centering
\includegraphics[width=0.7\columnwidth]{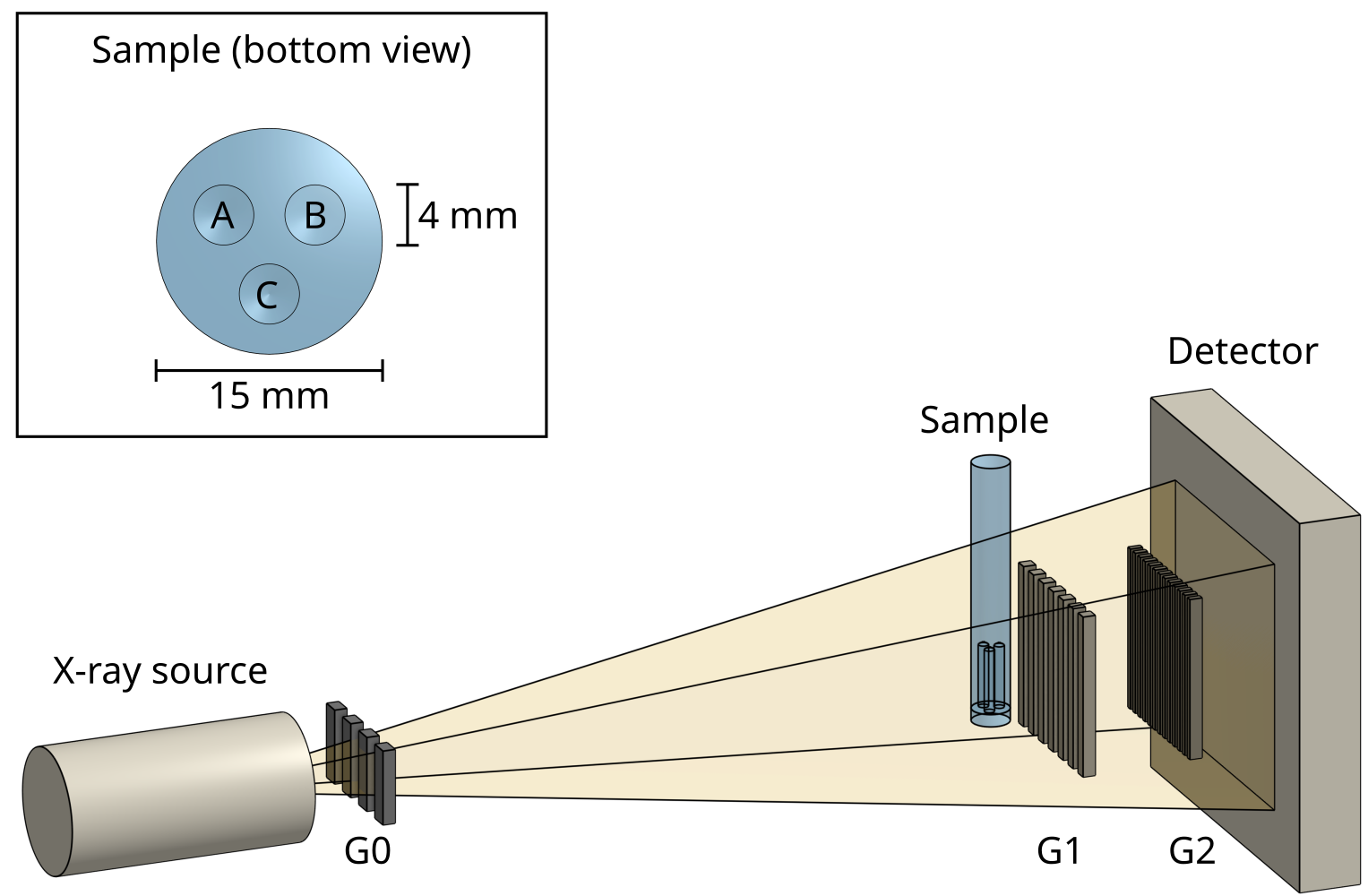}
\caption{Schematic of the Talbot--Lau interferometry setup and phantom used in the experiments. The vertical cavities (A, B, and C) are filled with sugar--water solutions of 10, 20 and 30 m-\%. The water container surrounding the sample is omitted for clarity.}
\label{fig:real:setup}
\end{figure}

Measurements were acquired at the University of Helsinki with the TLI setup of \cite{makinen2024optimization} (Fig.~\ref{fig:real:setup}) and are openly available \cite{makinen2026dataset}. The gratings G0, G1, and G2, have vertical line structures with periods $(14,3.5,2.0)$\,{\textmu}m, heights $(50,36,22)$\,{\textmu}m, and duty cycles $(0.3,0.5,0.5)$. The G0 absorption grating provides spatial coherence at the tube window. G1 is $140$\,cm downstream and produces a $\pi$ phase shift at $28$\,keV. The G2 absorption grating, $20$\,cm beyond G1 at the fifth fractional Talbot distance, analyzes the interference pattern. The gratings were fabricated at the Paul Scherrer Institut on silicon using photolithography, wet etching, and gold electroplating for G0 and G2 \cite{david2007gratings}.

The source (XRD Glass Tube, Malvern Panalytical) has a stationary tungsten anode, a $0.4\times8$\,mm focal spot viewed at a $6^\circ$ take-off angle, giving an effective focal spot of $0.4$\,mm vertical by $0.8$\,mm horizontal, and a $0.3$\,mm beryllium window.



Object and reference scans (without the object) used $10$ phase steps over one G2 period at each of $361$ projection angles over a full rotation, with $18$\,s exposure per step at $35$\,kV and $40$\,mA, i.e.\ about $18$\,h of exposure per scan and about $40$\,h in total, including sample repositioning.

The phantom (Fig.~\ref{fig:real:setup}), also used in \cite{makinen2024optimization}, is a $15$\,mm-diameter PMMA rod with three $4$\,mm-diameter vertical cavities filled approximately with $10$, $20$, and $30$~m-\% sugar--water solutions. At $5$\,cm before G1, it was submerged in a rectangular water container to reduce phase wrapping and beam-hardening artifacts. With the container in the beam, the mean reference visibility was about $15\,\%$.


\subsubsection{Data Retrieval and Preprocessing}\label{sec:exp:real:preproc}

Slice reconstruction uses $843$ pixels of the central detector row. At each of the $361$ projection angles, the stepping intensity is modeled as ~\cite{bech2010quantitative}
\begin{equation}\label{eq:stepping}
I(\phi_k) \approx c_0\left[\,1 + v\cos(\phi_k + \psi)\,\right],
\qquad k = 1,\dots,K,
\end{equation}
where $\phi_k$ are the step positions and $c_0$, $v$, and $\psi$ are mean intensity, visibility, and modulation phase. Single-harmonic Fourier analysis estimates these parameters from the $K=10$ steps for object and reference scans, giving
\begin{equation}\label{eq:retrieval}
A = -\ln\frac{c_0^{\mathrm{obj}}}{c_0^{\mathrm{ref}}},
\qquad
C = \psi^{\mathrm{obj}} - \psi^{\mathrm{ref}},
\qquad
D = -\ln\frac{v^{\mathrm{obj}}}{v^{\mathrm{ref}}},
\end{equation}
with $C$ wrapped to $(-\pi,\pi]$. $A$ and $D$ are line integrals, while $C$ is the derivative along the detector of the line integral of $\delta$, matching the forward model of Section~\ref{subsec:forward}.


Discarding one outermost detector bin leaves $842$ bins. For each projection and channel, the median of the $50$ sample-free bins at either edge is subtracted to remove the detector baseline; the sign of $A$ is set to the forward-model convention. FBP obtains integrated phase from $C$ by Fourier integration, using the same background bins to fix the integration constant. Bayesian reconstruction uses $C$ directly.


Intensity noise $\sigma_I$ is estimated from reference scan stepping-curve content outside the zeroth (DC) and first harmonics. Monte Carlo propagation through the complete retrieval pipeline uses $50$ noisy realizations of the object and reference scans; empirical standard deviations give $\sigma_A$, $\sigma_C$, and $\sigma_D$, including nonlinear noise amplification at finite visibility. Noise propagation is repeated for the five-step retrieval, giving increases of $1.413$, $1.416$, and $1.416$ in $\sigma_A$, $\sigma_C$, and $\sigma_D$, respectively, approximately $\sqrt{2}$. Each acquisition condition uses the noise levels estimated for its step count, held fixed during reconstruction.


Full data use $361$ angles and $10$ phase steps. Sparse-angle data use $30$ evenly spaced projections shared across channels; because both the $0^\circ$ and $360^\circ$ views are included, they cover $29$ distinct angles. Sparse-step data use $5$ evenly spaced steps spanning one grating period; combined undersampling applies both reductions. The three subsets contain $12$, $2$, and $24$ times fewer object exposures than the full scan ($3610$ exposures, $18.1$\,h at $18$\,s each); at fixed tube settings, exposure time and dose fall by the same factors.

\subsubsection{Reconstruction and Evaluation Settings}\label{sec:exp:real:settings}

The fan-beam model uses source-to-object and source-to-detector distances of $1.35$ and $1.63$\,m, $842$ detector bins of $74.8$\,{\textmu}m pitch, and a calibrated detector-center offset of four pixels. Reconstruction uses a $512^2$ grid over $52.2$\,mm ($102$\,{\textmu}m pixels), $842$ quadrature points per ray, and a nearest-neighbor projector; FBP uses a Ram--Lak filter. Bayesian inference is restricted to a circular support of normalized radius $0.5$ ($13.1$\,mm), outside which the images are fixed at zero; the support contains the rod and a margin of the surrounding water, with $51\,288$ active pixels per channel. The likelihood uses $(A,C,D)$ and the estimated noise levels.

The priors use channel scales \(s_c\) equal to half the range between the 5th and 95th percentiles of each channel's full-data FBP image, so the channel and coupling weights and the smoothing parameter $\beta$ are not directly comparable with the synthetic values. Prior parameters are selected as described in Section~\ref{sec:exp:real:params}. MAP estimates for both priors use zero-initialized L-BFGS with a strong-Wolfe line search and at most $150$ iterations. Posterior sampling uses the NUTS settings of the synthetic study (Section~\ref{sec:exp:synthetic:settings}), two chains of $1000$ samples, but with $200$ warmup iterations, and is performed on the scaled images $\boldsymbol{\mu}/s_\mu$, $\boldsymbol{\delta}/s_\delta$, and $\boldsymbol{\epsilon}/s_\epsilon$. The sampling dimension is $153\,864$.

Without ground truth, we assess noise-normalized sinogram residuals and visual agreement with the full-data reconstructions, and compare posterior means and standard deviations between priors.

\subsubsection{Forward-Model Adequacy}\label{sec:exp:real:validation}

We assess the channel models \eqref{eq:attenuation_forward}--\eqref{eq:darkfield_forward} and calibrated geometry by a round trip through the measured intensity, retrieval, and reconstruction pipeline.

\begin{figure}[!t]
\centering
 \includegraphics[width=0.9\columnwidth]{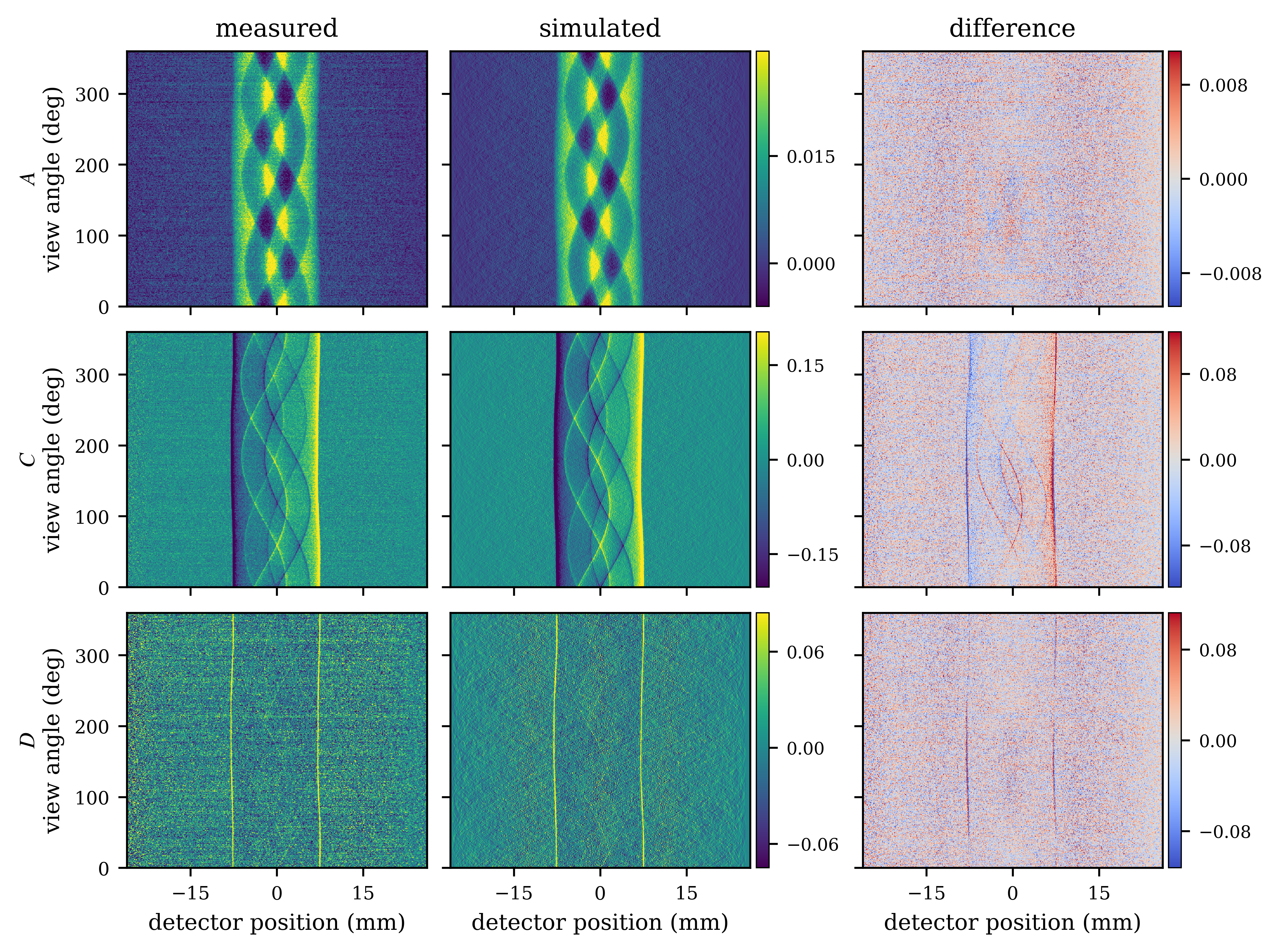}
\caption{Round-trip comparison of the full-data experimental sinograms. Rows: attenuation $A$, differential phase $C$, and dark-field $D$; columns: retrieved from the measured intensities, retrieved from the simulated intensities, and simulated minus measured. Within each row the measured and simulated panels share a color scale; the difference panel has its own, centered on zero.}
\label{fig:real:validation:sinos}
\end{figure}

Full-data FBP of the retrieved attenuation, integrated-phase, and dark-field sinograms gives $(\boldsymbol{\mu}_{\mathrm F},\boldsymbol{\delta}_{\mathrm F},\boldsymbol{\epsilon}_{\mathrm F})$. Inverting \eqref{eq:retrieval} generates simulated object intensities using the measured reference Fourier coefficients $\hat c_0=c_0$ and $\hat c_1=\tfrac12 c_0v e^{\mathrm{i}\psi}$:
\begin{equation}\label{eq:roundtrip}
\hat{c}_0^{\mathrm{sim}}
   = \hat{c}_0\, e^{-\mathcal{F}_\mu(\boldsymbol{\mu}_{\mathrm{F}})},
\qquad
\hat{c}_1^{\mathrm{sim}}
   = \hat{c}_1\,
     e^{-\mathcal{F}_\mu(\boldsymbol{\mu}_{\mathrm{F}})
        -\mathcal{F}_\epsilon(\boldsymbol{\epsilon}_{\mathrm{F}})}\,
     e^{\,\mathrm{i}\,\mathcal{F}_\delta(\boldsymbol{\delta}_{\mathrm{F}})},
\end{equation}
evaluated entry-wise. Measured reference coefficients retain the instrument response, including source spectrum, grating imperfections, and beam-profile variations, so the test isolates the object-dependent model. The simulated intensities pass through the same retrieval and FBP pipeline.
We compare measured and simulated intensities, sinograms (Fig.~\ref{fig:real:validation:sinos}), and FBP images using the relative $\ell_2$ discrepancy, normalized by the measured quantity.

The intensity discrepancy is $0.022$. Independent noise of equal size in the measured and simulated sinograms would give a nominal noise-only discrepancy of $\sqrt{2}\,n$, where $n$ is $\sigma_A$, $\sigma_C$, or $\sigma_D$ (Section~\ref{sec:exp:real:preproc}) relative to the RMS of the measured sinogram.  For $(A,C,D)$, discrepancies $(0.44,0.69,1.19)$ against $n = (0.28,0.42,0.76)$ are approximately $10\,\%$, $15\,\%$, and $11\,\%$ above this reference. Residual structure is concentrated near strong gradients, consistent with differences in projector and detector-derivative discretization.

The phase-image discrepancy is $0.25$, with SSIM $0.68$ and an approximately uniform offset of $15\,\%$ inside the rod. Attenuation and dark-field FBP comparisons are noise-dominated, with discrepancies of $1.31$ and $1.40$, close to $\sqrt{2}$. The round trip checks consistency among measured data, retrieval, geometry, and channel models; it does not establish the FBP images as ground truth.

\subsubsection{Prior Parameter Selection}
\label{sec:exp:real:params}

Without ground truth, channel weights are selected by held-out prediction: every fifth angle is withheld ($73$ of $361$), reconstructions use the remaining $288$, and candidates are scored by the held-out error, the relative $\ell_2$ error of the predicted held-out sinograms. This angular-stride split follows \cite{hendriksen2020}. The score does not involve the noise levels, but the reconstructions, and hence their ranking, depend on them through the likelihood; they are held fixed throughout the sweep.

We select $\beta$ and $\lambda_0$ on sparse-angle data using SSIM against full-data FBP as an imperfect reference. Only the phase channel is sensitive to $\beta$, with SSIM peaking at $\beta=0.03$. We use $\lambda_0=10^{-2}$, which adds curvature in weakly constrained directions; relative to $10^{-6}$, it moves only the phase MAP, $4\,\%$ further from FBP in relative $\ell_2$ with unchanged SSIM.

A logarithmic grid of $29$ channel weights from $10^{-4}$ to $10^3$ gives $\alpha_\mu=5.62$ and $\alpha_\epsilon=1.78$. Phase held-out error is minimal at $\alpha_\delta=17.78$ and within $0.42\,\%$ of that over $[10,31.6]$; we take the lower end, $\alpha_\delta=10$. This reduces the local curvature scale of the smoothed TV term ($\alpha_\delta/\beta$ from $593$ to $333$) at negligible cost in held-out error; in otherwise identical sparse-angle runs of the channel-wise posterior, it raises the median per-pixel phase ESS from $38.9$ to $99.9$. This choice also reflects the synthetic study’s sensitivity to excessive phase regularization with undersampled data (Section~\ref{sec:exp:synthetic:settings}).

Held-out error varies by less than $0.1\,\%$ over $\alpha_\mathrm J\in[0,2.15]$. We therefore compare sparse-angle JTV reconstructions with the full-data channel-wise TV reconstruction: mean SSIM is nearly constant for $10^{-4}\le\alpha_\mathrm J\le1$ and decreases for larger values, so we take $\alpha_\mathrm J=1$. This full-data reference is a surrogate, and each parameter group is selected with the others fixed, which need not be jointly optimal.

\subsubsection{MAP Reconstruction Results}\label{sec:exp:real:map}

\begin{figure*}[!t]
\centering
\includegraphics[width=0.9\textwidth]{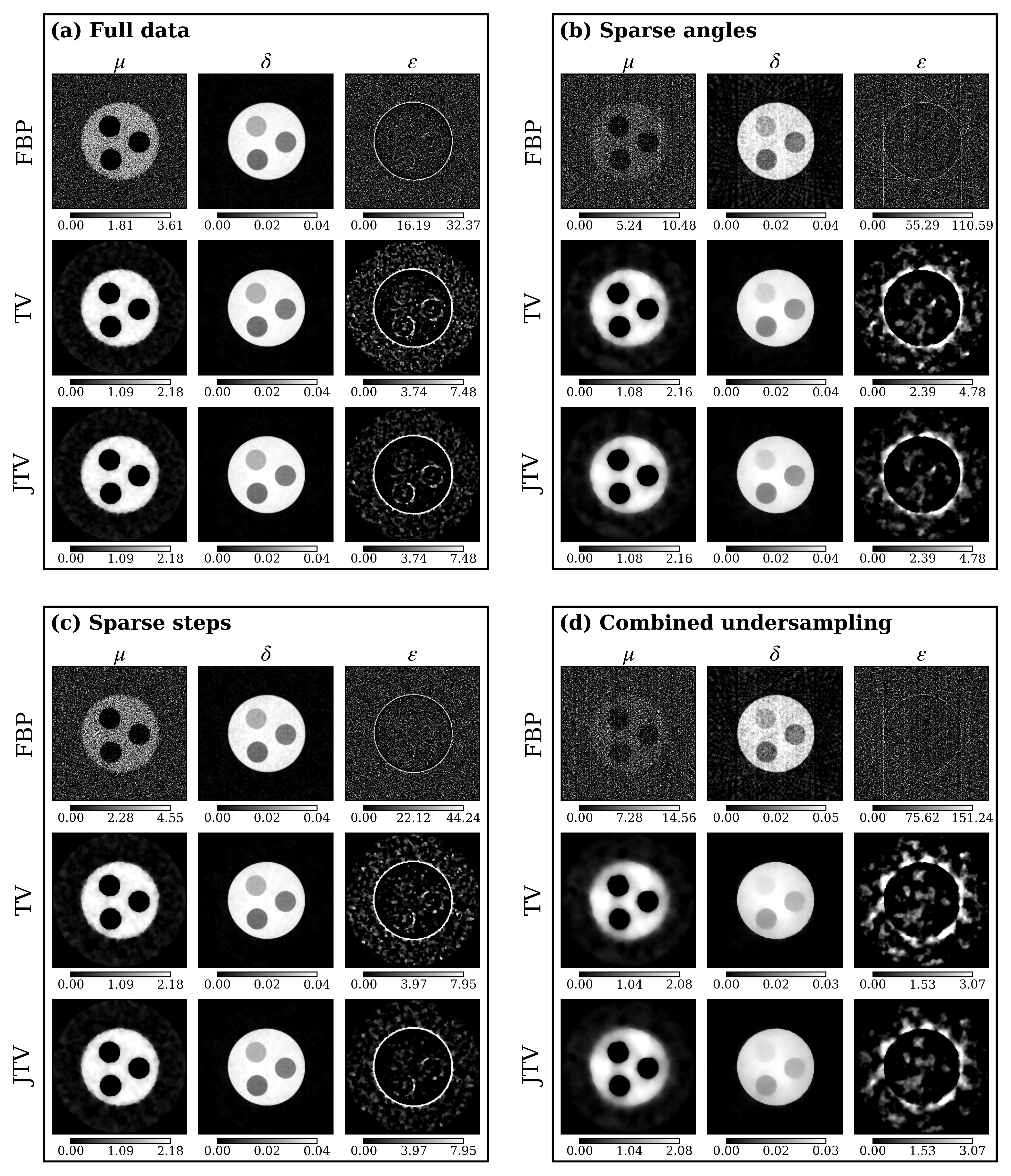}
\caption{Real-data reconstructions of attenuation ($\mu$), phase ($\delta$), and dark-field ($\epsilon$): (a) full data, 361 angles and 10 phase steps; (b) sparse angles, 30 angles and 10 steps; (c) sparse steps, 361 angles and 5 steps; (d) combined undersampling, 30 angles and 5 steps. Rows show FBP, TV, and JTV. Within each panel, TV and JTV share a scale per channel; FBP uses its own. Scales differ between panels. All panels are cropped to the bounding box of the reconstruction support, $26.1$\,mm across: TV and JTV are identically zero outside it, whereas FBP, which uses no support constraint, extends beyond the crop.}
\label{fig:real:sparse:map}
\end{figure*}

Figure~\ref{fig:real:sparse:map} shows severe sparse-angle FBP streaks, especially in phase, and noise-dominated dark-field images. TV and JTV suppress these artifacts and resolve all three cavities in $\mu$ and $\delta$. Their attenuation and phase images are similar; JTV's clearest effect is in the dark-field background outside the rod, where its standard deviation is $21$--$39\,\%$ lower than with TV. The dark-field rod boundary is also more continuous. MAP takes about $7$\,s for $30$ views and $40$--$50$\,s for $361$ views; FBP takes \mbox{$3$--$11$\,ms}.

Sinogram residuals, the RMS misfit of each channel in units of its noise standard deviation (Table~\ref{tab:real_datafit}), decrease from $1.43$--$4.06$ for FBP to $1.21$--$1.31$ for TV and JTV. Because each step count uses its own noise estimate, the residuals are comparable across conditions. TV gives smaller residuals throughout, with JTV values up to $1.4\,\%$ higher. These residuals measure data consistency, not reconstruction accuracy; values above unity indicate that the residual exceeds the estimated noise scale, potentially reflecting prior-induced bias, model discrepancy, or limitations of the noise model.

\begin{table}[!t]
\centering
\caption{Per-channel sinogram residual: the RMS misfit between predicted and measured sinograms in units of the channel's noise standard deviation, estimated for each step count. A value of $1$ means a residual at that noise level.}
\label{tab:real_datafit}

\footnotesize
\setlength{\tabcolsep}{5pt}
\renewcommand{\arraystretch}{0.95}

\begin{tabular}{cc cccc}
\toprule
Channel & Method & Full & Sparse & Sparse & Combined \\
        &        & data & angles & steps   &          \\
\midrule
\multirow{3}{*}{$\mu$}
& FBP & 1.553 & 4.064 & 1.529 & 4.020 \\
& TV  & 1.248 & 1.242 & 1.225 & 1.243 \\
& JTV & 1.249 & 1.248 & 1.226 & 1.252 \\
\cmidrule(lr){2-6}
\multirow{3}{*}{$\delta$}
& FBP & 1.471 & 4.040 & 1.428 & 3.477 \\
& TV  & 1.209 & 1.227 & 1.217 & 1.294 \\
& JTV & 1.211 & 1.241 & 1.220 & 1.311 \\
\cmidrule(lr){2-6}
\multirow{3}{*}{$\epsilon$}
& FBP & 1.559 & 3.987 & 1.538 & 3.977 \\
& TV  & 1.224 & 1.251 & 1.223 & 1.266 \\
& JTV & 1.235 & 1.268 & 1.232 & 1.273 \\
\bottomrule
\end{tabular}
\end{table}

\subsubsection{Posterior Sampling and UQ}
\label{sec:exp:real:uq}

Using the settings in Section~\ref{sec:exp:real:settings} and the prior parameters in Section~\ref{sec:exp:real:params}, the channel-wise and joint posteriors are sampled under all four conditions with the step-specific noise estimates. Adapted step sizes range from $3.8\times10^{-2}$ to $5.4\times10^{-2}$, with mean acceptance probabilities of $0.82$--$0.89$. A chain takes approximately $9.5$\,h for $361$ views and $50$\,min for $30$ views.

Diagnostics cover every sampled pixel. Full-data median ESS values are high (Table~\ref{tab:real_ess}), with median $\hat R=1.000$ in each channel. With sparse steps, median ESS falls by $19$--$41\,\%$ across channels and priors; maximum $\hat R$ remains below $1.07$. The larger noise variance weakens the likelihood relative to the fixed prior and changes posterior geometry, but these diagnostics alone do not establish the cause of slower mixing.


Angular undersampling particularly affects phase: TV median ESS falls to $99.9$ for sparse angles and $56.3$ for combined undersampling. For the channel-wise phase posterior, median $\hat R$ reaches $1.007$--$1.011$ and the $99$th percentile of $\hat R$ reaches $1.12$--$1.18$, with spatially coherent elevations and minimum ESS values of $2.7$ and $3.0$, respectively. The $99$th percentile of $\hat R$ remains below $1.06$ in $\mu$ and $\epsilon$. Although the medians and the $\mu$ and $\epsilon$ diagnostics are favorable, these tail values show unresolved slow mixing in parts of the phase posterior. The detector derivative in \eqref{eq:dpc_forward} suppresses low detector frequencies, so the data constrain slowly varying phase components only weakly; this plausibly contributes to the difficult phase conditioning, although the diagnostics do not identify the slowly mixing components directly.


JTV raises median ESS in every limited-data channel--condition pair by $4$--$38\,\%$. Under sparse angles, the phase $99$th percentile of $\hat R$ decreases from $1.12$ to $1.08$, its maximum from $1.33$ to $1.17$, and minimum ESS increases from $2.7$ to $5.1$. Under combined undersampling, the phase $99$th percentile of $\hat R$ is essentially unchanged at $1.18$, while median ESS rises only from $56.3$ to $58.7$, without improvement in its minimum. These gains are conditional on the different target posteriors and leave slow phase mixing unresolved.

\begin{table}[!t]
\centering
\caption{Median per-pixel ESS over the reconstruction support for the real-data posteriors, using two chains of $1000$ retained draws.}
\label{tab:real_ess}
\footnotesize
\setlength{\tabcolsep}{5pt}
\renewcommand{\arraystretch}{0.95}
\begin{tabular}{llrrr}
\toprule
Scenario & Prior & $\mu$ & $\delta$ & $\epsilon$ \\
\midrule
Full data     & TV  & 801.5 & 819.5 & 1043.6 \\
              & JTV & 867.6 & 889.7 & 1137.6 \\
Sparse angles & TV  & 186.9 & 99.9  & 209.5 \\
              & JTV & 224.9 & 114.2 & 289.0 \\
Sparse steps  & TV  & 547.2 & 486.3 & 778.6 \\
              & JTV & 657.6 & 580.0 & 927.0 \\
Combined      & TV  & 157.9 & 56.3  & 215.7 \\
              & JTV & 174.6 & 58.7  & 256.3 \\
\bottomrule
\end{tabular}
\end{table}

\begin{figure*}[!t]
    \centering
    \includegraphics[width=0.85\textwidth]{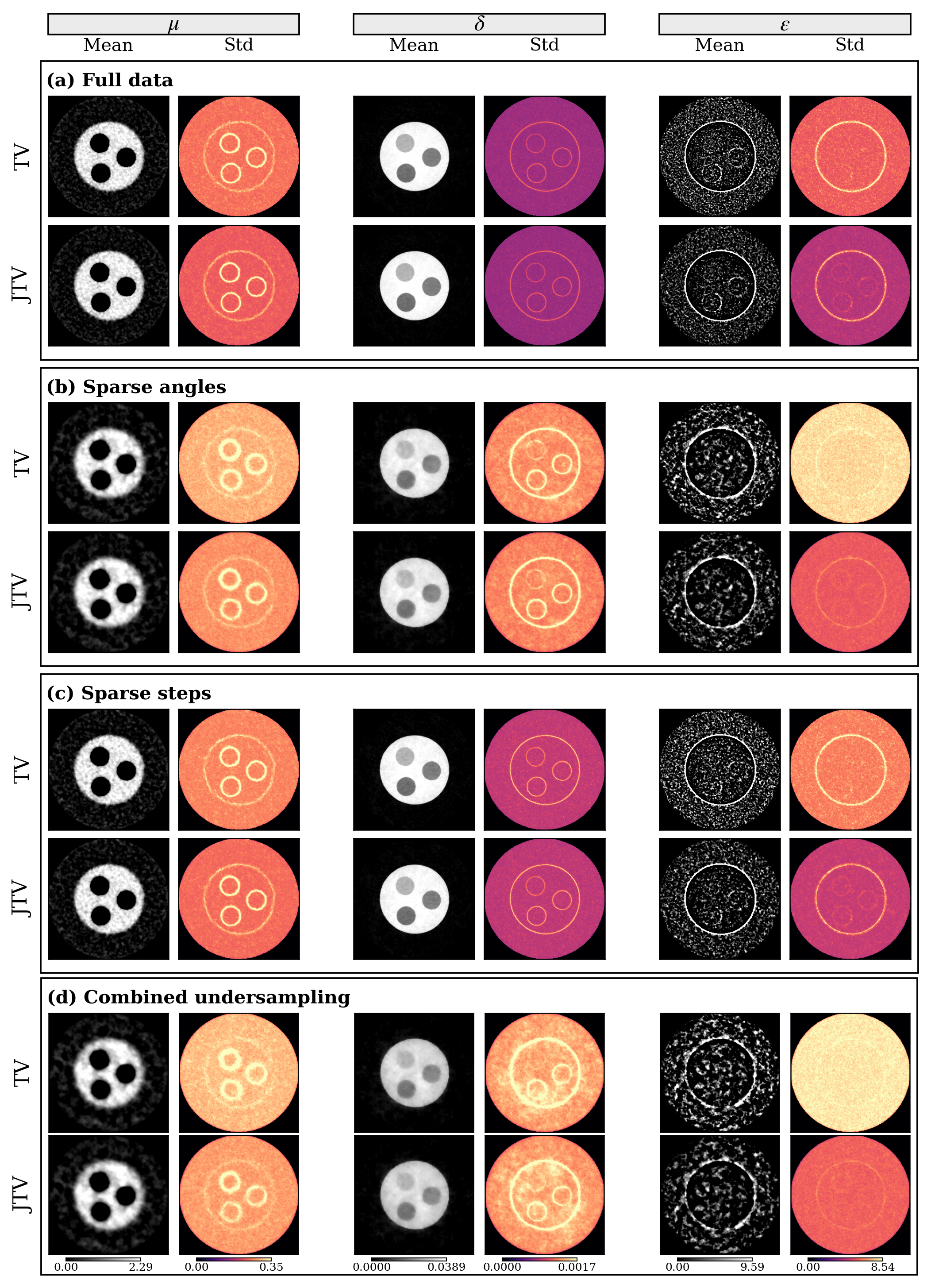}
    \caption{Real-data posterior means and pixel-wise standard deviations for TV and JTV under (a) full data, (b) sparse angles, (c) sparse steps, and (d) combined undersampling. Each quantity shares a scale per channel across all scenarios and priors; the bottom colorbars apply to every panel. All panels are cropped to the bounding box of the reconstruction support, $26.1$\,mm across, outside which the posterior is identically zero.}
    \label{fig:uq_comparison_real}
\end{figure*}

Figure~\ref{fig:uq_comparison_real} shows posterior means and pixel-wise standard deviations. Posterior-mean residuals range from $1.12$ to $1.24$, below the MAP residuals of Table~\ref{tab:real_datafit} in every case. Averaged over the reconstruction support, JTV lowers the standard deviation by $7$--$8\,\%$ in $\mu$, $1$--$3\,\%$ in $\delta$, and $23$--$30\,\%$ in $\epsilon$. For these acquisitions, reducing angular sampling has a larger effect on posterior spread than halving the phase steps. Relative to the full-data channel-wise posterior, halving the steps raises mean posterior standard deviation by $6\,\%$ in $\mu$, $20\,\%$ in $\delta$, and $12\,\%$ in $\epsilon$. Reducing $361$ views to $30$ raises it by $18\,\%$, $76\,\%$, and $38\,\%$, respectively, and combined undersampling by $21\,\%$, $89\,\%$, and $42\,\%$. The joint posterior follows the same pattern, with smaller dark-field increases of $26\,\%$ for sparse angles and $30\,\%$ for combined undersampling, each relative to its own full-data result. Phase is the most sensitive channel to both reductions, consistent with its derivative measurement. The posterior standard deviations are conditional on the assumed forward model, noise model, and prior, and carry additional Monte Carlo error where mixing is slow; the narrower JTV posteriors alone do not establish better calibration.

\section{Discussion}
\label{sec:discussion}


The joint formulation combines MAP estimation with posterior UQ for limited-data multi-contrast CT. Relative to channel-wise TV, JTV improves synthetic MAP accuracy modestly but consistently: RE decreases in $11$ of $12$ channel--condition pairs, and the summed per-channel RE by $0.8\,\%$ on average. Its clearest benefit is in the experimental dark-field channel, where it reduces the background standard deviation of the MAP image by $21$--$39\,\%$ and the mean posterior standard deviation by $23$--$30\,\%$, while MAP data-fit residuals increase slightly in all channels. These reductions in image noise and posterior spread do not by themselves imply improved reconstruction accuracy or calibrated uncertainty.


The synthetic results expose this distinction. In flat regions, RMS error is $0.37$--$0.74$ times RMS posterior standard deviation in $\mu$ and $\epsilon$, and $0.79$--$1.02$ times in $\delta$, whereas at boundaries the ratio reaches $2.5$--$3.2$ in $\mu$, $1.2$--$1.8$ in $\epsilon$, and $5.3$--$10.5$ in $\delta$ (Table~\ref{tab:synth:uq}). These spatial summaries from one noise realization are not repeated-sampling coverage estimates. Reconstructing phase with the simulation projector reduces edge error by about $40\,\%$, supporting a contribution from model discrepancy, while residual error can also reflect grid representation and prior-induced bias. The phase ratio decreases as data are removed because posterior spread increases while edge error persists. Thus, the standard deviation maps describe posterior variability under the assumed model and prior but can substantially understate boundary error.


The posterior mean was less accurate than the MAP estimate for $\mu$ and $\epsilon$ in every synthetic condition, most strongly for dark-field (RE $0.160$ versus $0.115$ with TV on full data). The weights were tuned for MAP accuracy, so this is an empirical comparison for the evaluated phantom, not a general ordering of Bayesian estimators. Parameter selection also matters: weights tuned to full synthetic data can over-regularize undersampled phase, whereas tuning across acquisition conditions keeps RE within $4\,\%$ of each per-condition optimum. Ground-truth tuning on the evaluated noise realization gives best-case synthetic performance; the experimental selection combines held-out prediction with surrogate full-data references and is sequential. Hierarchical estimation of the weights and noise levels \cite{bardsley2012mcmc}, or empirical Bayesian estimation of the weights \cite{vidal2020maximum}, could replace this selection, at additional sampling cost.


Operating after signal retrieval makes the framework transferable beyond TLI, for example to EI, when separately retrieved attenuation, differential phase, and dark-field sinograms are available and the forward operators and noise model are adapted. This modularity has a statistical cost: independent additive Gaussian noise in the retrieved channels approximates the nonlinear propagation of intensity noise and neglects the dependence of the noise level on transmission and visibility \cite{revol2010noise} and possible correlations; where strong scattering saturates the dark-field signal, retrieval itself becomes unreliable \cite{makinen2026adaptive}. The experimental residuals exceed the estimated noise scale, and the round-trip discrepancies are approximately $10$--$15\,\%$ above the nominal independent-noise reference (Section~\ref{sec:exp:real:validation}). These checks indicate remaining disagreement, without uniquely identifying its source. Intensity-domain inference could propagate uncertainty through retrieval within a joint model, but would require a more demanding nonlinear likelihood and would still need to address forward-model error.


At fixed tube settings, the sparse-angle, sparse-step, and combined experimental subsets correspond to $12$-, $2$-, and $24$-fold reductions in exposure time and dose (Section~\ref{sec:exp:real:preproc}), while the largest increase in channel-wise mean posterior standard deviation is $76$, $20$, and $89\,\%$, respectively, each in phase. 
These retrospective estimates of potential savings apply to the water-immersed phantom, without dosimetry or validation at equal image quality. The water container increases exposure requirements \cite{makinen2024optimization}; performance on heterogeneous specimens scanned in air remains untested, particularly when phase wrapping violates the assumed observation model.

Sampling cost and convergence limit quantitative UQ. A $361$-view experimental chain takes about $9.5$\,h for $153\,864$ unknowns. JTV raises median ESS by $4$--$38\,\%$ in the limited-data runs and lowers the sparse-angle phase $99$th percentile of $\hat R$ from $1.12$ to $1.08$. Slow phase mixing persists, however, and under combined undersampling the phase $99$th percentile of $\hat R$ remains approximately $1.18$ for both priors. The smoothed TV term has its largest local curvature in flat regions, where it scales as the channel weight divided by $\beta$; reducing $\alpha_\delta/\beta$ from $593$ to $333$ raised the median per-pixel ESS of the sparse-angle channel-wise phase posterior by a factor of $2.6$, while that of $\mu$ and $\epsilon$ changed by at most $9\,\%$. Within each geometry, the four experimental posteriors took the same time per iteration to within $0.3\,\%$, consistent with trajectories reaching the tree-depth limit in every run. Five-step retrieval approximately doubles the noise variance, halving the likelihood's weight relative to the fixed prior, while median ESS falls by $19$--$41\,\%$ relative to full data across channels and priors. These observations suggest sensitivity to the balance between likelihood and prior, although posterior geometry and sampler adaptation also affect mixing. Further validation should quantify the Monte Carlo error of the reported uncertainty summaries with longer or additional chains and assess their stability across noise realizations.

Extension to three dimensions will require more scalable inference. Candidate approaches include whitened latent Gaussian fields, such as Whittle--Mat\'ern models represented by truncated Karhunen--Lo\`eve expansions, combined with nonlinear maps that represent sharp interfaces \cite{yazdanian2026bayesian}; channel coupling could then act on the latent fields or on the reconstructed images. Intensity-domain noise models, repeated-sampling coverage studies, and further biomedical and industrial datasets are complementary priorities.

\section{Conclusion}
\label{sec:conclusion}


We studied joint Bayesian reconstruction of attenuation, phase, and dark-field images from retrieved sinograms under limited-data acquisition. A JTV prior provides structural coupling within MAP estimation and posterior sampling. Across synthetic and experimental TLI studies, MAP reconstruction with either prior suppresses undersampling artifacts relative to FBP, while JTV gives modest synthetic accuracy gains and a cleaner experimental dark-field background than channel-wise TV.



For the experimental data, JTV lowers mean dark-field posterior standard deviation by $23$--$30\,\%$ and raises median per-pixel ESS by $4$--$38\,\%$ under limited data. These gains describe the posterior under the assumed model and sampling efficiency, not calibration: synthetic posterior spread matched or exceeded the error in flat regions but understated it at boundaries, and slow phase mixing persists under angular undersampling. Future work will address uncertainty calibration, intensity-domain likelihoods, and scalable inference for three-dimensional multi-contrast CT.

\section*{Acknowledgment}
The authors used ChatGPT (OpenAI) and Claude (Anthropic) for language editing and polishing, and Claude Code for assistance with software development and repository organization. The authors reviewed and edited the assisted material and take full responsibility for the final manuscript, code, and results.

\bibliographystyle{ieeetr}
\bibliography{references}

\vfill

\end{document}